\documentclass[preprint,review,12pt]{elsarticle}

\usepackage{graphicx}
\usepackage{amssymb}
\usepackage{lineno}

\newcommand{\rmi}{\mathrm{i}}

\journal{Physica C}

\begin{document}

\begin{frontmatter}



\title{Directional pinning and anisotropy in YBa$_2$Cu$_3$O$_{7-x}$ with BaZrO$_3$ nanorods: intrinsic and nanorods-induced anisotropy}


\author[RomaTre]{N. Pompeo}
\author[RomaTre]{K. Torokhtii}
\author[ENEA]{A. Augieri}
\author[ENEA]{G. Celentano}
\author[ENEA]{V. Galluzzi}
%
%\author[ENEA]{eneidi}
%
\author[RomaTre]{E. Silva\corref{cor1}}
\cortext[cor1]{Corresponding author}
\ead{enrico.silva@uniroma3.it}
\address[RomaTre]{Dipartimento di Ingegneria and Unit\`a CNISM,
Universit\`a Roma Tre, Via della Vasca Navale 84, 00146 Roma,
Italy}

\address[ENEA]{ENEA-Frascati, Via Enrico Fermi 45, 00044 Frascati, Roma, Italy}

\begin{abstract}
We present a study of the anisotropic vortex parameters as obtained from measurements of the microwave complex resistivity in the vortex state with a tilted applied magnetic field in YBa$_2$Cu$_3$O$_{7-x}$ thin films with BaZrO$_3$ nanorods. We present the angular dependence of the vortex viscosity $\eta$, the pinning constant $k_p$ and the upper limit for the creep factor $\chi_M$. We show that the directional effect of the nanorods is absent in $\eta$, which is dictated by the mass anisotropy $\gamma$. By contrast, pinning-mediated properties are strongly affected by the nanorods. It is significant that the pinning and creep affected by the nanorods is detectable also at our very high operating frequency, which implies very short-range displacements of the vortices from their equilibrium position.

\end{abstract}


\end{frontmatter}

% \linenumbers

% main text
\section{Introduction and model}
\label{intro}
It is nowadays established that BaZrO$_3$ (BZO) inclusions in YBa$_2$Cu$_3$O$_{7-x}$ (YBCO) films provoke a strong enhancement of the pinning properties \cite{macmanusNATMAT04,kangSCI06,gutierrezNATMAT07}. In pulsed-laser-deposited (PLD) films the BZO inclusions can self-assemble as nanorods approximately perpendicular to the $a,b$ planes \cite{augieriIEEE09}, thus introducing a source of anisotropy in addition to the $a,b$ planes themselves. Disentangling the effect of those two sources of anisotropy is not trivial: the quasi-layered structure determines both a mass anisotropy in the electronic properties, $\gamma^2=m_c/m_{ab}$, as well as directional pinning along the planes. BZO determines basically directional pinning along the nanorods, i.e. parallel to the $c$ axis of the films. BZO pinning extends for tens of degrees away from the nanorod direction \cite{zuevAPL08,pompeoAPL13}, thus mixing up with the narrow-angular-range $a,b$ plane pinning. While this effect is beneficial for applications, because it increases the apparent isotropic pinning, it is an obstacle to the understanding of the different roles of the different sources of anisotropy.

Aim of this paper is to present a small-perturbation study of the vortex dynamics with the perspective of clarifying the different roles of the mass anisotropy and of the two sources of directional pinning. To this aim, we measured the very high frequency (48 GHz) complex response of PLD YBCO films with BZO nanorods, as a function of the dc magnetic field orientation $\theta$ with respect to the $c$ axis. The high frequency induces very small oscillations of the vortex matter (less than $\sim 1$ nm \cite{tomaschPRB88}), and no plastic deformation of the vortex system is determined by the applied stimulus (differently, e.g., from critical current density $J_c$ measurements). From the complex microwave response we extracted \cite{pompeoPRB08} the relevant vortex parameters: the vortex viscosity $\eta$, the pinning constant (Labusch parameter) $k_p$, and the upper limit for the creep factor $\chi_M$. We discuss in this paper the angular dependence of all the vortex parameters.

The guiding model for the interpretation of our data is a very general mean-field, relaxational model for vortex dynamics. The basic equation for the complex vortex motion resistivity reads \cite{pompeoPRB08}:
\begin{equation}
\label{eq:rhovm}
    \rho_{vm}=\left(\rho_{v1}+\rmi\rho_{v2}\right)f_L(\theta)=\mu_0H\frac{\Phi_0}{\eta}\frac{\chi+\rmi\left(\nu/\bar{\nu}\right)}{1+\rmi\left(\nu/\bar{\nu}\right)}f_L(\theta)
\end{equation}
\noindent where the vortex viscosity is related to the genuine $a,b$ plane flux-flow resistivity through $\rho_{ff,ab}=\Phi_0\mu_0H/\eta$, where $\Phi_0$ is the flux quantum and $\mu_0$ the free space permeability. All vortex parameters are, in principle, field- and angle- dependent, and $f_L(\theta)$ explicitly includes the possible variation of the Lorentz force acting on flux lines when the field is tilted \cite{pompeoCM13} (this point will be specifically discussed in the next Section). The characteristic frequency $\bar{\nu}$  contains also the dimensionless parameter $0\leq\chi\leq 1$, which in turn is a measure of thermal activation phenomena. Only when $\chi\rightarrow 0$, $\bar\nu$ coincides with the depinning frequency $\nu_p=k_p/2\pi\eta$. 
The relations of Eq.(\ref{eq:rhovm}) with several models \cite{GR,CC,brandtPRL91} have been extensively discussed elsewhere \cite{pompeoPRB08}, together with the applicability limits and the systematic errors in the extraction of the vortex parameters, so that we recall here only the results: (i) in the measurements of the angular and field dependence of $\eta$ and $k_p$  here reported, the systematic (not scattering) error is less than 20\%; (ii) it is possible to obtain, from the same set of data, the upper limit for the creep factor, $\chi_M$. Related studies have been presented in \cite{related}. We mention that to obtain the numerical values of $k_p$  the adoption of a specific model is required, although the differences between analyses performed with different models are practically limited to scale factors. When needed, we adopted here the  Coffey-Clem model \cite{CC}.
\section{Experimental results}
\label{Exp}
%
%\noindent
%
\subsection{Sample and experimental methods}
\label{sample}
We focus here on an YBCO/BZO sample grown by PLD from targets containing BaZrO$_3$ (BZO) powders at 5\% mol. \cite{galluzziIEEE07}, which showed very strong BZO-enhanced pinning even at microwave frequencies \cite{pompeoAPL07} and it has been extensively characterized. TEM images demonstrated that BZO determines elongated, columnar-like defects, perpendicular to the film plane \cite{augieriJAP10}. Consistently, the angular dependence of the critical current density exhibits a peak when the field is perpendicular to the film plane \cite{augieriIEEE09}. 

A dc magnetic field $H$ was applied at the angle $\theta$ with the $c$ axis. We performed a field sweep in perpendicular orientation, i.e. for $\theta=0^{\circ}$, and rotations at fixed fields $H_i=$0.4 T, 0.6 T. The temperature was fixed at $T=80$~K as a compromise to obtain a reasonable signal for all orientations. We use the method of surface perturbation of a dielectric resonator working at 48 GHz \cite{pompeoJS07}. Measurements of the quality factor $Q$ and the resonant frequency $f_0$ as a function of the applied field H and the field orientation $\theta$ yield 
the vortex motion complex resistivity $\rho_v=\rho_{v1}+\mathrm{i}\rho_{v2}$ through the relation:
\begin{eqnarray}
\label{eq:exp}
  G\left[ \frac{1}{Q(H,\theta)}-\frac{1}{Q(0)} - 2\mathrm{i} \frac{f_0(H,\theta)-f_0(0)}{f_0(0)}\right]\simeq\\
  \nonumber
  \frac{\rho_{v1}(H,\theta)+\rm{i}\rho_{v2}(H,\theta)}{d}f_L(\theta)
\end{eqnarray}
\noindent where the zero-field contribution has been subtracted out, $G$ is a geometrical factor, the approximate equality holds since the thickness $d$ of our film is smaller than the penetration depth \cite{silvaSUST96}, and the Lorentz force contribution $f_L(\theta)$ takes into account the fact that our microwave setup induces circular, $a,b$ plane microwave currents in the film. $f_L(\theta)$ has been calculated as \cite{pompeoCM13}:
\begin{equation}
\label{eq:fteta}
    f_L(\theta)=\frac{\frac{1}{2}\gamma^{-2}\sin^2\theta+\cos^2\theta}{\gamma^{-2}\sin^2\theta+\cos^2\theta}
\end{equation}
where $\gamma=$5 in our sample \cite{pompeoAPL13}.
\subsection{Raw data and evidence for extrinsic effects}
\label{raw}
\begin{figure}[h]
\centerline{\includegraphics[width=7.5cm]{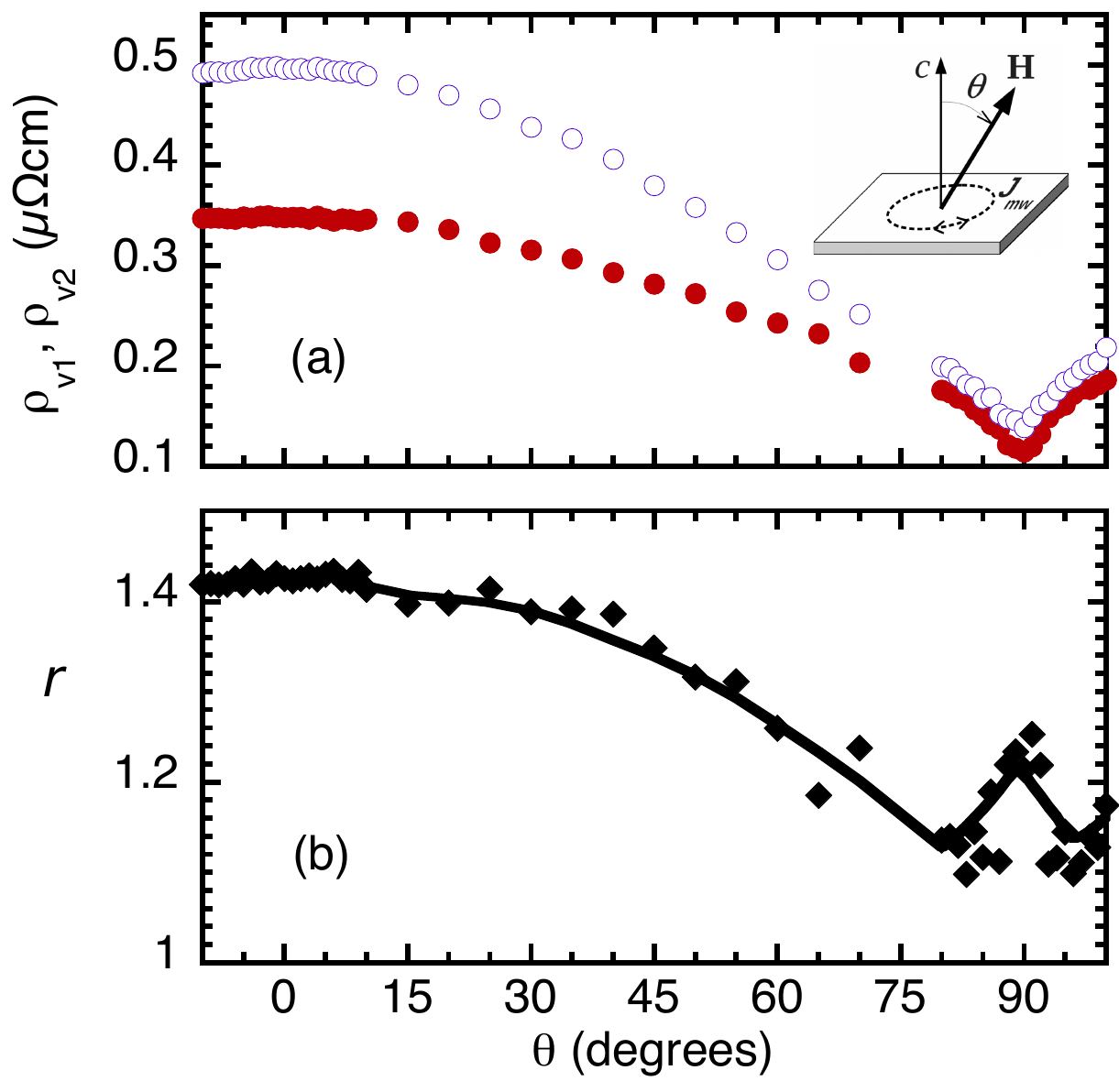}}
%\vspace{-10mm}
  \caption{(a) Angular dependence of the real (full red symbols) and imaginary (open blue symbols) vortex motion resistivity at $T=80$ K and $\mu_0 H = 0.4$ T. Inset: microwave current pattern and the definition of the angle $\theta$. (b) Angular dependence of the pinning parameter $r=\rho_{v2}/\rho_{v1}$ at $\mu_0 H_i=$ 0.4 T. Black symbols: experimental data. The continuous line is a guide to the eye.}
\label{figraw}
\end{figure}
It is interesting to discuss first the raw data, to stress the complex angular dependence of the electrical response in these nanostructured materials. Figure \ref{figraw} reports as an example the angular dependence of $\rho_{v1}, \rho_{v2}$ at $\mu_0H=$ 0.4 T. The dependence is clearly very regular, with a steady decrease approaching $\theta=0^{\circ}$. One may be tempted to assign this behavior to the mass anisotropy (and, to a lesser extent, to the reduced Lorentz force in our configuration), thus neglecting the role of nanorod-induced pinning. However, the complexity of the system is revealed by the analysis of the $r$ parameter \cite{halbritterJS95}, defined as $r=\rho_{v2} / \rho_{v1}$. The $r$ parameter is an experimental measure of the relative weight between reactive and dissipative contributions to the response and, as such, is an indicator for strong ($r>1$) or weak ($r<1$) pinning. In the present case, it should be noted that $r$ is obtained directly from the raw data (see Eq.(\ref{eq:exp})), and thus  it is not affected by the evaluation of $f_L(\theta)$. The behavior of $r(\theta)$ reported in Figure \ref{figraw} points to strong pinning in {\em perpendicular} orientation ($\theta = 0^{\circ}$), a crossover to dissipative response toward parallel orientation, and a sharp enhancement of pinning when $\theta = 90^{\circ}$. Thus, it is clear that nanorods play a major role in the vortex response even for very small oscillations of the vortices.

A second indication of the relevance of directional pinning is the failure of the angular scaling, according to the well-established scaling rules \cite{BGL,HC}: when only point pinning is present, or no pinning is effective, a physical observable in an anisotropic superconductor should depend on the field $H$ and angle $\theta$ only through $H\varepsilon(\theta)$, with $\varepsilon(\theta)=\left[\cos^2\theta+\gamma^{-2}\sin^2\theta\right]^{1/2}$ in the anisotropic 3D Ginzburg-Landau model. An example of successful scaling is the dc resistivity in the pinning-free region \cite{sartiPRB97}. Thus, the quantities $\rho_{v1}(H_i,\theta), \rho_{v2}(H_i,\theta)$ should recover the field-sweep $\rho_{v1}(H,0^{\circ}), \rho_{v2}(H,0^{\circ})$ if their anisotropy is dictated by the effective mass alone.
\begin{figure}[h]
\centerline{\includegraphics[width=7.5cm]{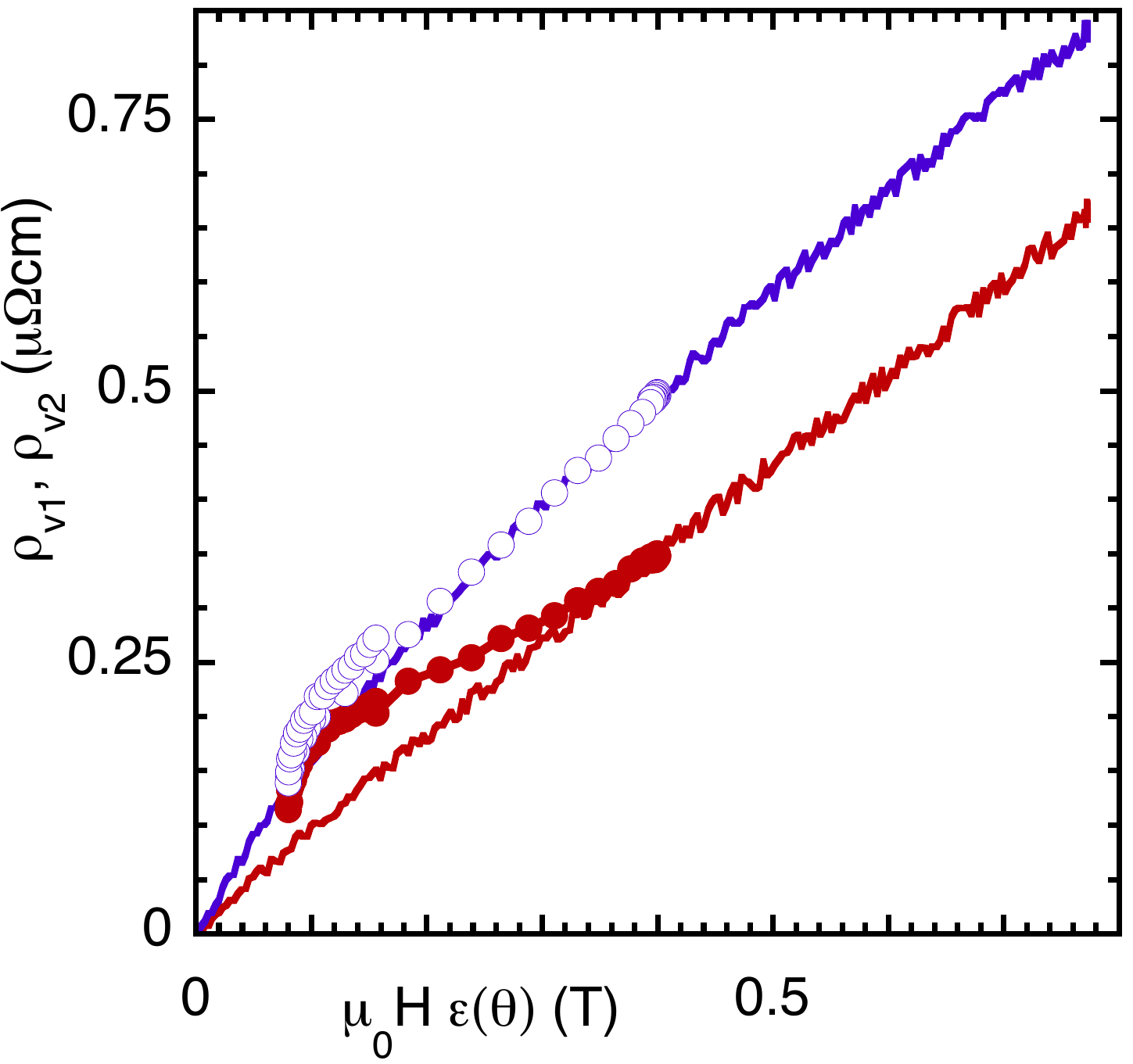}}
%\vspace{-10mm}
  \caption{
Tentative scaling of the data in Figure \ref{figraw}a over the field sweep taken in perpendicular orientation ($\theta=0^\circ$), by plotting the data as a function of $H\varepsilon(\theta)$ with $\gamma=5$. Symbols as in Figure \ref{figraw}a, continuous lines are the field sweep (red: $\rho_{v1}$, blue: $\rho_{v2}$). 
It is readily seen that the angular scaling for the raw data is impossible: only part of the data scale, limited to the imaginary part.}
\label{fignoscaling}
\end{figure}

Figure \ref{fignoscaling} reports the tentative scaling for $H_i=0.4$ T. It is immediately seen that one can scale, e.g., $\rho_{v1}/d$, but not simultaneously both components \cite{pompeoJS13}. Thus, this is a direct evidence that extrinsic and directional effects are present. Clearly, in view also of the results in Figure \ref{figraw}, the obvious  candidate is directional pinning. 
 
 A quantitative discussion of the vortex parameters is thus in order.

\subsection{Vortex viscosity}
\label{eta}

\begin{figure}[h]
\centerline{\includegraphics[width=7.5cm]{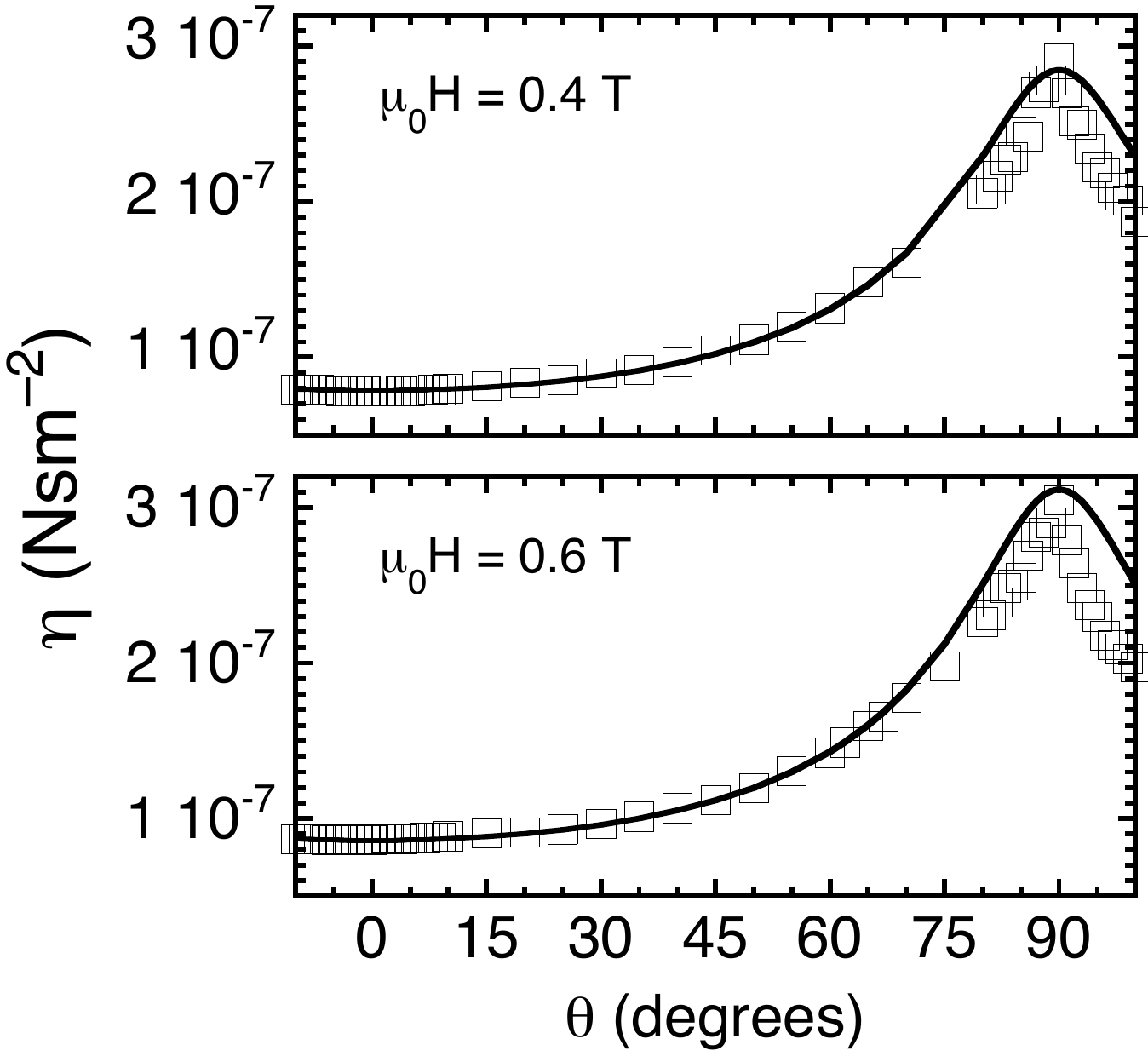}}
%\vspace{-10mm}
  \caption{Angular dependence of $\eta$ as derived from the angular data at $\mu_0H_i=0.4$ T, 0.6 T (upper and lower panel, respectively). The continuous lines are fits with $\eta(\theta) = \eta(0^{\circ})\varepsilon^{-0.8}(\theta)$ (see text), with $\gamma=5$.}
\label{figeta}
\end{figure}

The vortex viscosity, as a parametrization of the flux flow resistivity, is a property that should not depend on pinning. In the simple Bardeen-Stephen model \cite{BS} one has $\rho_{ff,ab}\propto H/H_{c2}(\theta)=H\varepsilon(\theta)$, where $H_{c2}=H_{c2}(0^{\circ})/\varepsilon(\theta)$ is the upper critical field, so that  $\eta(\theta) \propto \varepsilon^{-1}(\theta)$. Thus, no features due to BZO directional pinning along the $c$ axis should be detected. Figure \ref{figeta} shows that this is indeed true. In our specific case, if we take into account (see Figure \ref{fignoscaling}) that the complex resistivity is sublinear, we find \cite{pompeoAPL13} $\rho_{ff,ab}\propto (H\varepsilon(\theta))^{0.8}$, whence $\eta(\theta) = \eta(0^{\circ})\varepsilon^{-0.8}(\theta)$, in very good agreement with the data taking $\gamma=5$ (Figure \ref{figeta}).
\subsection{Pinning constant and creep factor}
\label{kp}
\begin{figure}[h]
\centerline{\includegraphics[width=7.5cm]{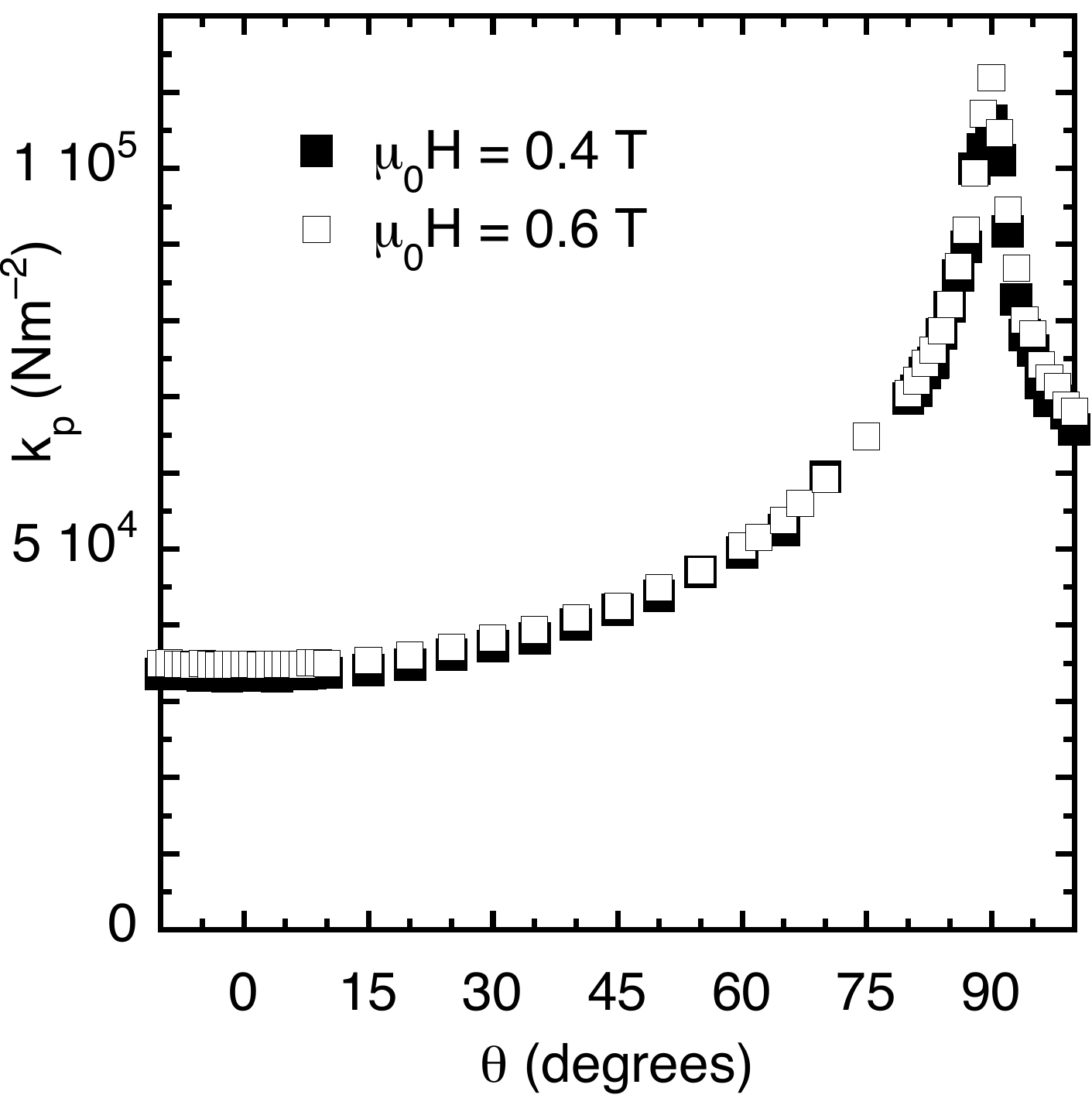}}
%\vspace{-10mm}
  \caption{Angular dependence of $k_p$ as derived from the angular data at $\mu_0H_i=$0.4 T, 0.6 T (full and open squares, respectively). The two sets of data superimpose, thus indicating no field dependence of $k_p$ in this field range.}
\label{figkp}
\end{figure}

The pinning constant $k_p$ is reported in Figure \ref{figkp}. A preliminary analysis clearly show that the apparent anisotropy of $k_p$ is not related to the effective masses in a simple way. First, the anisotropy is too low: $k_p(90^{\circ})/k_p(0^{\circ})\simeq 3$, as opposed to $\gamma=5$ as found in the previous analysis and in a more refined treatment \cite{pompeoAPL13}. It might be justified on the basis of a field dependence $k_p\propto (B\varepsilon(\theta))^{\alpha}$ with a properly chosen $\alpha$, but as reported in Figure \ref{figkp} no significant field dependence appears. Finally, the angular dependence itself is much steeper than a reasonable power of $\varepsilon(\theta)$. Thus, the effective anisotropy of $k_p$ is unlikely to be dictated by the effective mass anisotropy.

A thorough discussion of the role of the BZO-induced pinning anisotropy, with a comparison with angular measurements of $J_c$, has been reported previously \cite{pompeoAPL13}, and we do not repeat it here. Instead, we complete the discussion of the vortex parameters with the creep-related properties.

\begin{figure}[h]
\centerline{\includegraphics[width=7.5cm]{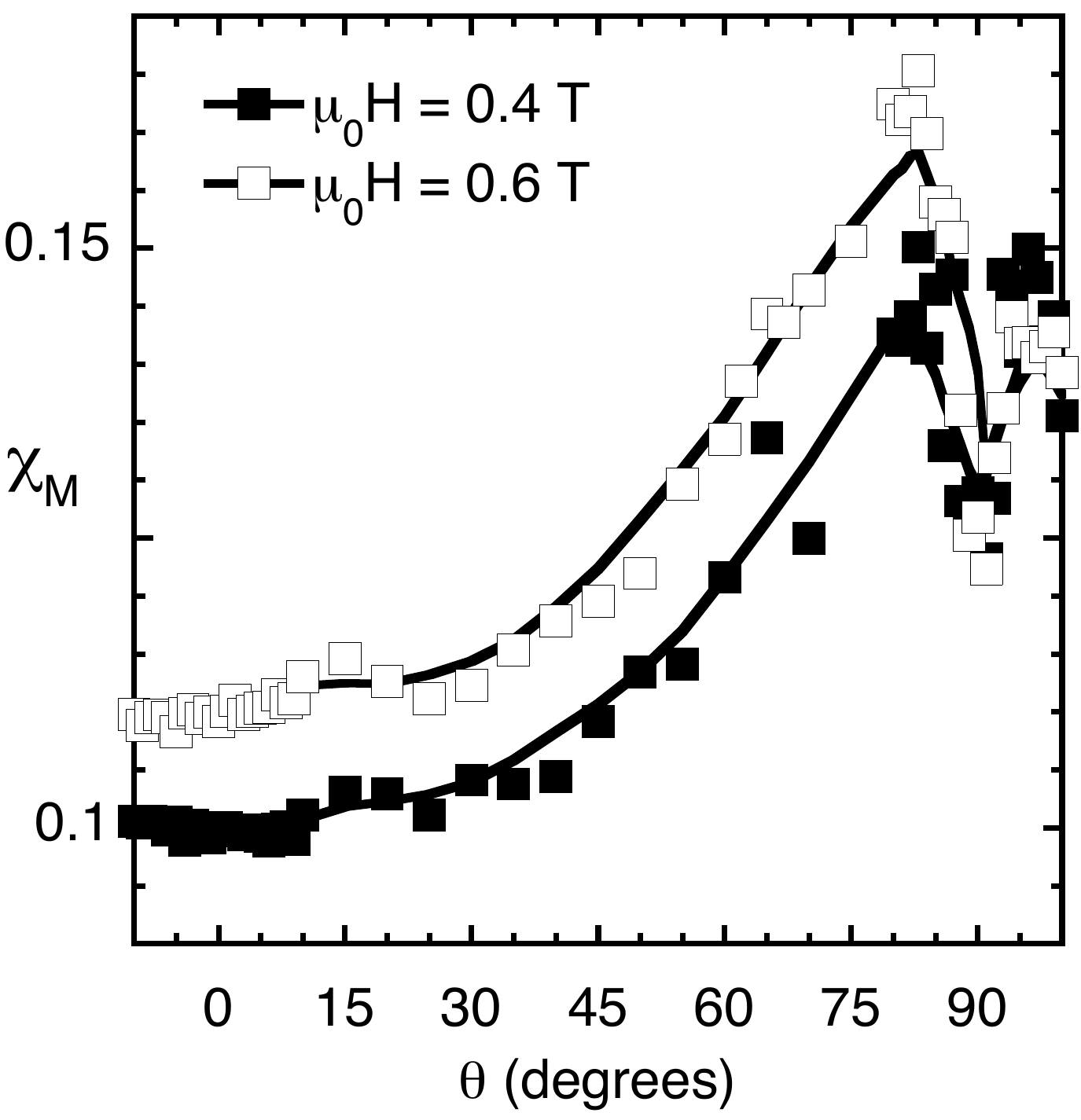}}
%\vspace{-10mm}
  \caption{Angular dependence of the maximum creep factor $\chi_M$ as derived from the angular data at $\mu_0H_i=$0.4 T, 0.6 T (full and open squares, respectively). The low values of $\chi_M$ along the directional pinning due to BZO are evident. Continuous lines are guides to the eye.}
\label{figcreep}
\end{figure}

The data allow for the extraction of the upper limit of the creep factor, $\chi_M$. The creep factor is a direct measure of the weight of flux creep: $\chi=0$ means no creep at all, while $\chi=1$ indicates free flux lines. In Figure \ref{figcreep} we report the angular dependence of the maximum creep factor $\chi_M$. As it can be seen, $\chi_M < 0.18$ in the full angular range, pointing to deep pinning wells. The angular dependence clearly indicates that creep is unfavored especially in the region where BZO nanorods are efficient, and in a narrow angular range around the $a,b$ planes. While $k_p$ is directly related to the curvature of the pinning well at the bottom (due to the small oscillation regime here probed), $\chi$ is affected also by the depth. Even if our data allow only the extraction of an upper limit for $\chi$, we believe that the indication of deep pinning wells along the BZO nanorods is clear enough. We mention that this effect is mainly related to single-vortex dynamics: when dc measurements (such as $J_c$) are examined, quantitative differences emerge when the field approaches the nanorods direction, that can be explained in terms of the different  dynamics and the possible vortex-Mott-insulator phase \cite{pompeoAPL13}.
\section{Concluding remarks}
\label{conc}
From the analysis of the microwave data, it is clear that in YBCO/BZO the anisotropic behavior of the electrical response in the vortex state is very complex. In particular, the interplay between the directional effect of BZO nanorods, the layered structure and the mass anisotropy can be unveiled when more than a single observable is measured. From measurements of the complex resistivity one has access to both the vortex viscosity $\eta$, dominated by the mass anisotropy $\gamma$ and by the angular function $\varepsilon(\theta)$, and the pinning constant $k_p$, dominated by the specific anisotropic pinning mechanisms. In addition, the upper limit for the creep factor add information on the extrinsic (defect-dominated) properties, and it stress the relevance of the directional pinning determined by BZO nanorods.
\\
\section*{Acknowledgements}
\label{ack}
This work has been partially supported by the FIRB project ``SURE:ARTYST" and by EURATOM. N.P. acknowledges support from Regione Lazio.

%% References


%\bibliographystyle{elsarticle-num}



\end{document}